\documentclass[twocolumn,superscriptaddress,10pt,aps,prb]{revtex4-2}
\usepackage{epsfig,amsopn}
\usepackage{graphicx}
\usepackage{epstopdf}
\usepackage{braket}
\usepackage{amsmath,amssymb}
\usepackage{amsthm}
\usepackage{enumerate}
\newcommand*{\balancecolsandclearpage}{%
	\close@column@grid
	\cleardoublepage
	\twocolumngrid
}

\newcommand\bs{\boldsymbol}

\newcommand\s{\sigma}

\usepackage[defaultcolor=blue]{changes}
\definechangesauthor[name=Rekha, color=blue]{rk}

\newcommand{\RNum}[1]{\uppercase\expandafter{\romannumeral #1\relax}}

\definechangesauthor[name=Arijit, color=magenta]{ak}

\begin{document}
	\title{Valley filtering and valley valves in irradiated pristine graphene}  
	\author{Rekha Kumari}
	\email{rekha@iitk.ac.in}
	\affiliation{Department of Physics, Indian Institute of Technology, Kanpur, India}
	\author{Gopal Dixit}
	\email{gdixit@phy.iitb.ac.in}
	\affiliation{Department of Physics, Indian Institute of Technology Bombay, Powai, Mumbai 400076, India}
	\author{Arijit Kundu}
	\email{kundua@iitk.ac.in}
	\affiliation{Department of Physics, Indian Institute of Technology, Kanpur, India}

\begin{abstract}
We theoretically study  valley-filtering in pristine graphene irradiated by bicircular counter-rotating laser drive. The dynamical symmetry of the graphene and laser drive disrupts graphene's inversion symmetry, which results  distinct quasi-energy states  and Floquet band occupations in the two valleys. 
Controlling the relative phase between the bicircular laser drive ultimately allows to blocks the contribution from one valley while allowing the opposite valley currents in  the system. 
For practical  realization of valley-based device, we propose configurational setup for valley filters and valley valve consisting of two graphene nanoribbons irradiated by two bicircular counter-rotating laser drives with a relative phase shift.  It is observed that the relative phase  between the two bicircular  laser drives offer a control knob to generate  valley-selective currents and  transport responses with very high efficiency by an all-optical way.  
In addition, our findings about valley filter and valley valve are robust against moderate disorder and modest changes in driving laser parameters.
Present work opens  an avenue to realise light-based valleytronics devices in reality. 
\end{abstract} 

\maketitle
	
\section*{Introduction}
Apart from spin, electrons in  two-dimensional (2D) materials possess an addiational quantum attribute, namely, 
valley degree of freedom~\cite{vitale2018valleytronics, bussolotti2018roadmap, Xiao_2007}. 
This valley attribute is associated with the minima of the energy landscape in the momentum space, and
holds tremendous potential for diverse applications in information processing, optoelectronic devices, and quantum technologies. 
Capitalising on this potential, the field of valleytronics has emerged, aiming to exploit the valley attribute of 2D materials  for numerous technological applications~\cite{Culcer_2012,Niklas_2012,Laird_2013, mak2018light, schaibley2016valleytronics}.

In inversion symmetry broken 2D materials, controlling the valley attribute is relatively straightforward by exploiting the valley dependence of the Berry curvature and orbital angular momentum~\cite{Yao_2008,Zhu_2011,Xiao_2012,Cao_2012}. Experimental demonstrations involving circularly polarized light  
and magnetic field have successfully manipulated the valleys in such 2D  materials, resulting in valley-dependent transport signatures~\cite{mak2012control,Zeng2012control,Sallen2012control,guddala2021control,Yilei2014contol,aivazian2015control,MacNeill2015control,srivastava2015control,Symm_cite1}. 
Valley-selective responses pose a significant challenge in inversion-symmetric 2D  materials, such as graphene, primarily due to vanishing Berry curvature~\cite{Novoselov_VSG_2004,novoselov_VSG_2005,zhang_VSGExp_2005}. Nonetheless, several proposal have been put forward in achieving valley dependent transport in these systems by breaking the inversion symmetry externally~\cite{rycerz2007_VSBG,Gunlycke2011_VSBG,Hunt2013_VSBG,Marko2014_VSBG,Settnes2016_VSBG, AK_2016,Faria2020_VSBG, shimazaki2015generation, tamura2023origins, tapar2023effectuating, liu2023valley, ortiz2022graphene, golub2014valley}.
	
The application of periodic drives has emerged as a promising approach to control the properties quantum systems~\cite{Kitagawa_2010,Rudner_2013,Platero_2013,Piskunow_2014,Fidkowski_2019,Oka_2009,Lindner_2011,Lababidi_2014,Piskunow_2015,Eckardt_2017,Morimoto_2017,Harper_2020,Molignini_2020,Zhang_2021,wang2023topological, nag2019dynamical, ikeda2023photocurrent,Biswas_2020,Kundu_2020,Mohan_2016,Kundu_2014}. In recent years it has been demonstrated that tailored laser pulses can effectively controlled and manipulated valley selective outputs in inversion-symmetry systems~\cite{mrudul2021light, rana2023all, mrudul2021controlling, rana2022generation, sharma2023giant, avetissian2023graphene, morimoto2022atomic, kelardeh2022ultrashort}.
Introduction of the bicircular counterrotating laser configuration with control over subcycle phase allowed valley-selective excitation in pristine graphene. In addition, controlling the laser parameters allows for precise manipulation of valley occupations, thus facilitating the generation of desired valley properties~\cite{mrudul2021light}.
Most of the recent works have focused on infinite 2D sheet of the pristine graphene to demonstrate valley-selective excitation. 
However, finite system size is required to conceive any practical valley-based device. 
In addition, disorder is unavoidable during the sample preparation. Thus, it is not {\it a priori} obvious how the finite size and the presence of the disorder affect  valley-related  properties in pristine-graphene based valleytronics device. 
These open questions are  serious  impediment in harnessing the full potential of inversion-symmetric 2D materials, 
and practical realization of valley-based devices for upcoming quantum technologies.
Present work addresses these crucial challenges and thus improve our understanding   in realising graphene-based valleytronics device closer to the reality. 

Present work demonstrates  the potential of configuring devices using an inversion-symmetric graphene, which can generate valley-selective outcomes in transport experiments. 
For this purpose, we introduce a device configuration that employs zigzag graphene nanoribbons subjected to bicircular laser fields, serving as a valley filter.
This  device configuration is capable of producing a controlled valley-selective output in two-terminal transport devices.  Furthermore, we illustrate  that the connection of two valley filters in series creates a valley valve device, highlighting its ability to realise ``perfect" valley valve in transport measurements. 

We employe Floquet non-equilibrium Green's function approach to simulate the outcomes   
of the two-terminal conductance and valley polarization for the valley filter and valley valve devices. 
The valley-selective outputs are qualified using  non-identical valley occupation of the quasi-energy states within Floquet framework. 
Additionally, to assess the experimental feasibility of these valley-selective effects, we have 
explored robustness of our findings in presence of static disorder.
	
\begin{figure*}[tb]
\centering
\includegraphics[width=.95\linewidth]{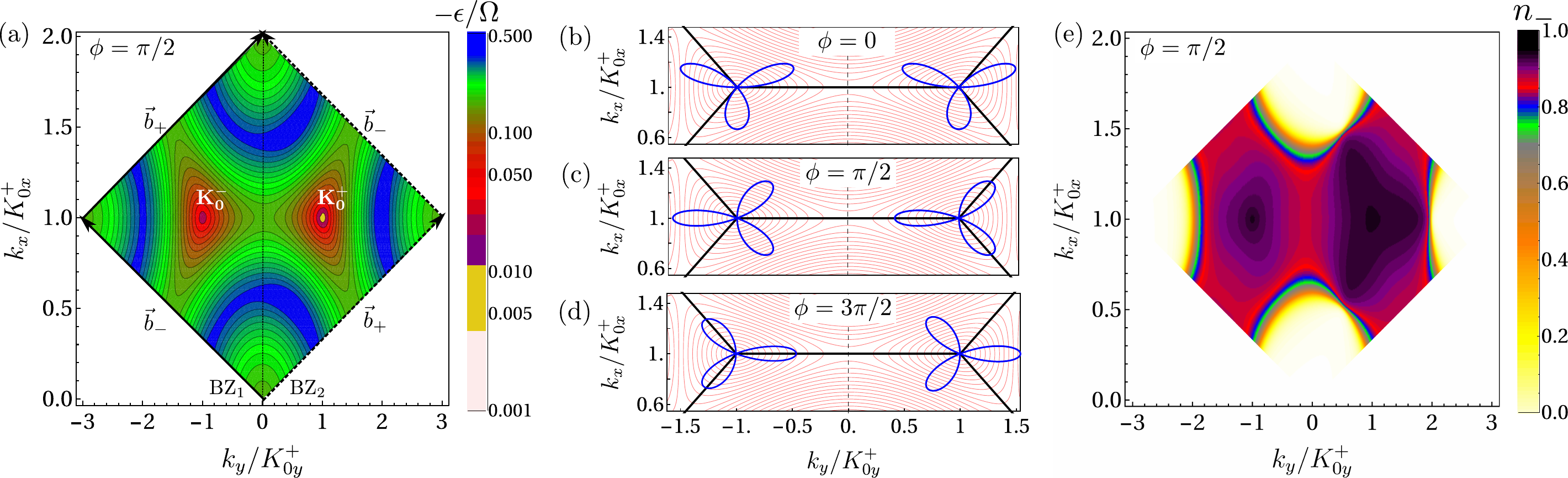}
\caption{(a) Quasi-energy  of the valence band as a function of crystal momentum $k_x/K^{+}_{0x}$ and $k_y/K^{+}_{0y}$ in units of $1/\Omega$. The solid and dashed vectors $\bs{b}_{\pm}$ represent the Brillouin zone of the two Dirac points marked by $K^{+}_{0}$ and $K^{-}_{0}$. Projection of the Lissajous profiles  associated with the total vector potential of the bicircular counter-rotating laser fields with $p=-2$ onto the low-energy contours of a pristine graphene for different values of the phase: (b) $\phi=0$, (c) $\phi=\pi/2$, and (d) $\phi=3\pi/2$. (e) Occupation of the quasi-energy states  in the Brillouin zone for $\phi=\pi/2$. 
The parameters for (b-d) are $\mathcal{A}_{x} = \mathcal{A}_{y}=0.35$, $T= 1$ and  $\mathcal{R} = 1$. Parameters for (a) and (e) are   $\mathcal{A}_{x} = \mathcal{A}_{y}=1/2$, $T= {\gamma_{0}}/4$  and $\mathcal{R}=1$.}
\label{fig:n2hbz}
\end{figure*}

\section{Theory}
Periodically driven graphene can be described by the following Hamiltonian as~\cite{reich2002tight}
\begin{equation}\label{eq:H1}
\mathcal{H}(\bs{k},t) = -\gamma_{0}
\begin{pmatrix}
			0 & h(\bs{k}, t)\\
			h{^*}(\bs{k}, t) & 0  \end{pmatrix}, 
\end{equation} 
where $h(\bs{k})=\sum_{j} e^{i\bs{k}(t) \cdot {\bs{\delta}_j}}$ is the hopping term with 
${\bs{\delta}_j}$ as the vector connecting a sub-lattice $A$ with  its nearest neighboring $B$ sub-lattices  and $\gamma_0$ is the nearest neighboring hopping strength, which is chosen as 2.7 eV~\cite{AK_2016, mrudul2021high}. The index $j$ runs over the $A$ sub-lattice points. In the presence of a laser field, crystal momentum $\bs{k}$ changes to $\bs{k}(t) = \bs{k} + e \mathcal{A}(t)$, where the electric charge $e$ is taken as unity and the total vector potential $\mathcal{A}$ of a bicircular laser fields consist of two circularly polarized lasers with frequencies $\Omega$ and $p\Omega$; and  is expressed as 
\begin{eqnarray}\label{eq:At}
 \mathcal{A}(t) & = &  \mathcal{A}_{x}\big[\cos(\Omega t)+\mathcal{R}\cos(p\Omega t+\phi)\big]\hat{e}_x  \\ \nonumber
		& & +\mathcal{A}_{y}\big[\sin(\Omega t)+\mathcal{R}\sin(p\Omega t+\phi)\big] \hat{e}_y .
\end{eqnarray}
Here, $\mathcal{R}=\mathcal{A}_{p}/{\mathcal{A}_0}$ is the strength of the second laser 
field with respect to the fundamental field, $\mathcal{A}_{0}=\sqrt{\mathcal{A}_{x}^2+\mathcal{A}_{y}^2}$ and $p$ is an integer. The additional parameter $\phi$ accounts phase difference between these two fields. 
	
The positions of the Dirac points in the Brillouin zone (BZ) of a pristine graphene  
are denoted by $K^{\pm}_0=\left(2\pi/3a_0, \pm2\pi/3\sqrt{3}a_0\right)$ with $a_0$ represents the distance between two nearest carbon atoms in graphene. We set $a_0=1$ as our unit of length. The Dirac points in graphene correspond to two distinct valleys situated at  $\bs{b}_{\pm}=\{2\pi/3a_0,\pm 2\pi/\sqrt{3} a_0\}$, which are shown by the shaded vectors in Fig.~\ref{fig:n2hbz}(a). 
In the absence of irradiation, we define the mark parts of the BZ corresponding to these two valleys, valley-1 and valley-2, as BZ$_1$ and BZ$_2$, respectively, which  can be mapped  to each other using sub-lattice inversion as $\s_x \mathcal{H}(\bs{k})\s_x = \mathcal{H}(-\bs{k})$, when $\mathcal{A}=0$.
	
\begin{figure*}[ht!]
\includegraphics[width=0.95\linewidth]{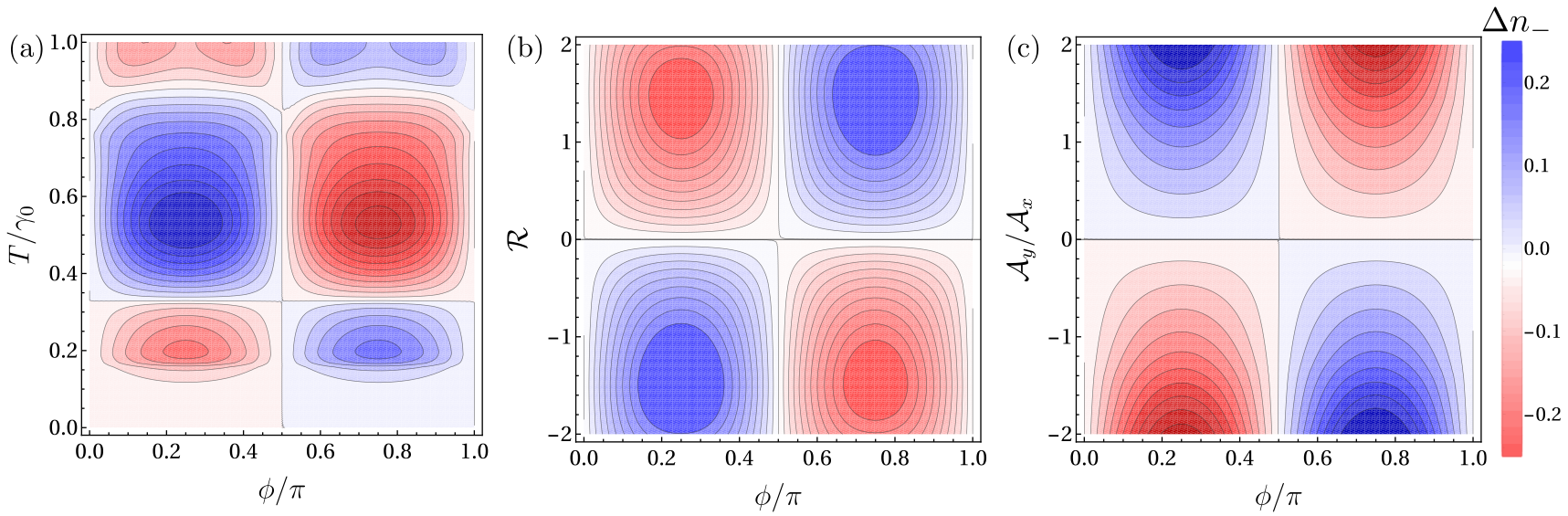}
\caption{Variations in the difference in the integrated valley occupations of the valence band  ($\Delta n_{-}$) 
as a function of (a) the time period ($T$),  (b) the ratio ($\mathcal{R}$) and (c) $\mathcal{A}_{y}/\mathcal{A}_{x}$ for different values of $\phi$. The integrated valley occupations are computed using eq.~(\ref{eq:doccu}) for $p=-2$ and $\mu=0$. The other parameters are $\mathcal{A}_{x} = \mathcal{A}_{y} = 1/2$ and $\mathcal{R} = 1$ for 
(a), $\mathcal{A}_{x} = \mathcal{A}_{y} =  1/2$ and $T=\gamma_0/4$ for  (b), and $\mathcal{A}_{x}=1/2, 
T=\gamma_0/4$, and $\mathcal{R}=1$ for  (c).}
\label{fig:MFIG_doccu_n2}
\end{figure*}
		
In the case of periodically driven  graphene, electron dynamics is determined by solving  time-dependent Schr{\"o}dinger equation as  $i\hbar\,\partial_t|\psi(t)\rangle = \mathcal{H}(\bs{k},t))|\psi(t)\rangle$ with $|\psi(t)\rangle$ as the time-dependent wave function. This time-dependent equation  has a complete set of solutions known as Floquet states, which  can be expressed as $|\psi_{\alpha}(t)\rangle = e^{-i\epsilon_{\alpha}t}|u_{\alpha}(t)\rangle$, where $|u_{\alpha}(t)\rangle$ is time-periodic part of the solution, i.e.,  $|u_{\alpha}(t)\rangle = |u_{\alpha}(t+T)\rangle$, and $
\epsilon_{\alpha}$ is the associated quasi-energy. 
These quasi-energies are unique only within a ``Floquet zone" defined by $-\Omega/2 \leq \epsilon_{\alpha} \leq \Omega/2$.
The effective Floquet Hamiltonian within  the sublattice basis, $(u_A(t), u_B(t))^T$,  can be expressed as
\begin{equation}
		\mathcal{H}_{\textrm{F}}(\bs{k},t)=
		\begin{pmatrix}
		-i\partial_t & -\gamma_0 h(\bs{k},t)\\
		-\gamma_0 h^{*}(\bs{k},t) & -i\partial_t
		\end{pmatrix}.
\end{equation}
It is known that the occupancy of the Floquet states deviates from the equilibrium distribution function in 
such a periodically driven system. 
Under the assumptions of the weak coupling to a fermionic reservoir,
 the occupations of a Floquet state with quasienergy $\epsilon_{\alpha}$ can be expressed as~{\cite{kumari2023josephsoncurrent,staircase_occupation}}:
\begin{equation}\label{eq:occu1}
		n_{\alpha}(\mu)=\sum_{l}f(\epsilon_{\alpha}+l\Omega-\mu)\langle u_{\alpha}^{(l)}|u_{\alpha}^{(l)}\rangle.
\end{equation}
Here, $\mu$ is the chemical potential of the reservoir and $f(x)=1/(e^{-\beta x}+1)$ is the Fermi distribution function with $\beta$ being the inverse of the temperature of the reservoir.

For the static system, the presence of an inversion symmetry  ($\bs{k} \rightarrow -\bs{k}$) connects the two Dirac points through lattice inversion, and ensures identical valley responses  in  pristine graphene. 
However, this situation changes drastically when graphene is expose to  the bicircular laser fields.
In this scenario, a form of optical inversion symmetry denoted as $\mathcal{P} = \sigma_x \left.\mathcal{K}\right|_{t \rightarrow -t}$ with $\mathcal{K}$ as the complex conjugation operator plays a similar role. 
This optical symmetry relates the two Dirac points as $\mathcal{P} \mathcal{H}_{\textrm{F}}(K_0^{+},t)\mathcal{P}^{-1} = \mathcal{H}_{\textrm{F}}(K_0^{-},t)$ for $\phi = n\pi$ with $n$ as an integer On the other hand, the optical valley symmetry is absent for $\phi \neq n\pi$, leading to the distinctive valley responses.
	
For the specific case of the bicircular laser fields with parameters $p = -2$ and $p = 4$, 
the effective Floquet Hamiltonian exhibits an additional temporal symmetry as
\begin{equation}
	\mathcal{H}_{\textrm{F}}(K_0^{\pm}, t + T/3) = e^{\mp 2i\pi/3}\mathcal{H}_{\textrm{F}}(K_0^{\pm}, t).
\end{equation}
This temporal symmetry is a consequence of the interplay between the temporal characteristics of the driving bicircular field 
and the inherent $\mathcal{C}_3$ rotational symmetry of graphene lattice. 
The application of the bicircular field results in the generation of Lissajous profiles that exhibit a distinctive trifold structure, which align perfectly with the valley-specific low-energy energy contours observed in pristine graphene. 
In addition, the orientation of these trifold structure can be  controlled by tuning the phase of the bicircular laser field.

The projection of the Lissajous profiles on the energy contours of pristine graphene is presented in   	
Fig.~\ref{fig:n2hbz}(b-d). The bicircular field interacts uniformly with both valleys for $\phi = 0$, and therefore preservers the valley symmetry as evident from Fig.~\ref{fig:n2hbz}(b).  In this case, the Lissajous profiles in both valleys and their corresponding low-energy spectra can be mapped onto each other through optical inversion, ensuring identical valley responses. This situation changes drastically  for $\phi = \pi/2$, where the bicircular field interacts differentially with the two valleys as illustrated by Fig.~\ref{fig:n2hbz}(c). 
In this case,  the Lissajous profile aligns perfectly with the low-energy contours of valley-1, while no such alignment occurs in the other valley. This situation can be reversed for $\phi = 3\pi/2,$ where the Lissajous profile aligns perfectly with the other valley as reflected from  Fig.~\ref{fig:n2hbz}(d). 
In these two later scenarios, the two valleys and their respective Lissajous figures cannot be mapped onto each other through optical inversion. The optical inversion symmetry breaking results in the emergence of valley polarization effects. The ability to control the alignments of these trifold patterns with individual valleys by varying $\phi$ of the bicircular field  offers a route to control valley-selective responses.

Momentum-resolved quasi-energy spectrum of an irradiated graphene for $\phi = \pi/2$ is shown in  	
Fig.~\ref{fig:n2hbz}(a). The absence of an inversion symmetry results in gapless states in one valley 
and a finite energy gap in the other valley as evident from the figure. 
This observation is complemented by analysing the momentum-resolved occupation of the quasi-energy states in Fig.~\ref{fig:n2hbz}(e), which is computed using Eq.~(\ref{eq:occu1}). 
The observed asymmetry in the quasi-energy spectrum between the two valleys leads to variations in the occupation of valley-specific quasi-energy states. These differences in occupation near the valleys give rise to valley asymmetric transports, which will be discussed  in detail later.

\begin{figure*}[th!]
\centering
\includegraphics[width=0.95\linewidth]{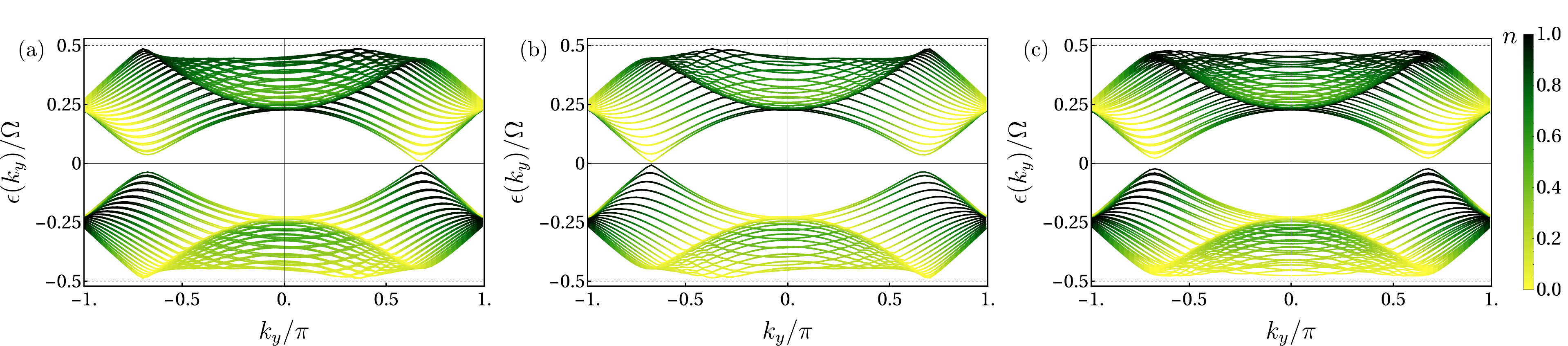}\vspace{-4mm}
\caption{Quasi-energy spectrum of a graphene nanoribbon of length $L = 40 a_{0}$ under bicircular laser fields with periodic boundary conditions as a function of crystal momentum $k_y$ (in units of $1/\pi$) for (a) $\phi=0$, (b) $\phi=\pi/2$, and (c) $\phi=3\pi/2$. 
The valley-specific regions of the one-dimensional BZ lie along positive and negative  $k_y$ axis.  
The colorbar represents the occupation of the quasi-energy states for $\mu=0$, i.e., charge neutrality. 
Rest of the parameters are the same as in Fig.~\ref{fig:n2hbz}(d) except $\mathcal{R} = 1.13$.}
\label{fig:qesdens}
\end{figure*}

\subsubsection*{Valley occupation differences} 
The difference in the occupation for the quasi-energy band (indexed by $\alpha=\pm$) between the two valleys can be computed as
\begin{equation}\label{eq:doccu}
	\Delta n_{\alpha} = \left[\int_{\rm{BZ}_1}n_{\alpha}(\bs{k}) - \int_{\rm{BZ}_2} n_{\alpha}(\bs{k}) \right] d\bs{k},
\end{equation}
where the integrals are performed by dividing the entire BZ  into two parts, BZ$_1$ and BZ$_2$, 
each containing position of one of the Dirac nodes ($K_{0}^{\pm}$ points) as discussed in Fig.~\ref{fig:n2hbz}(a). 
	
Figure~\ref{fig:MFIG_doccu_n2}(a) illustrates the difference in the integrated valence band occupation ($\Delta n_{-}$)
with respect to the time period of the irradiation ($T=2\pi/\Omega$) and $\phi$, while maintaining other  parameters constant,
namely $\mathcal{A}_{x} = \mathcal A_{y} =1/2$ (in unit where we set the electron charge $e=1$ and the lattice spacing $a_0=1$, as mentioned earlier) and $\mu=0$. 
The occupation differences exhibit periodic behavior  with a period of $\pi$, which 
can be comprehended through an examination of the symmetries inherent in Lissajous profiles and the $\mathcal{C}_3$ symmetry of the graphene lattice. 

Both  valleys  exhibit identical response for the integer multiples of $\pi$. Consequently, the integrated occupation differences amount to zero. 
However, the Lissajous profile aligns perfectly with the low-energy contours of one of the  valleys for $\phi=\pi/2$, giving rise to valley asymmetry and a non-zero valley response. 
Changing  $\phi$ from $\pi/2$  to $3\pi/2$ reverses the alignment and leads an interchange in the valley responses. 
Thus, the integrated occupation difference becomes the negative.
These integrated occupation differences provide the overall measure of the valley asymmetry.
Note that the occupation difference for the conduction band  is expressed as the negative of the 
integrated valley occupation difference of the valance band, and denoted as $\Delta n_{+}=-\Delta n_{-}$ for $\mu=0$. 

Sensitivity of the variations in occupation differences  as a function of the laser's parameters $\mathcal{R}$ and $\mathcal{A}_y/\mathcal{A}_x$ with $\phi$ are presented in Figs.~\ref{fig:MFIG_doccu_n2}(b) and (c), respectively.  
Beyond a critical value of $\mathcal{R}$, which exceeds 1, we observe the emergence of symmetrical trifold structures corresponding to two valleys at the center. 
A similar trend is evident in the integrated valley occupation difference as depicted in Fig.~\ref{fig:MFIG_doccu_n2}(b). 
As $\mathcal{R}$ increases, $\Delta n_{-}$ rises until it approaches 1, and further increments in $\mathcal{R}$ lead to a reduction in $\Delta n_{-}$. Figure~\ref{fig:MFIG_doccu_n2}(c) shows an increase in the ratio $\mathcal{A}_y/\mathcal{A}_x$ results in a corresponding increment in the integrated valley occupation differences.  
Changing the sign of these ratios, $\mathcal{R}$ and  $\mathcal{A}_y/\mathcal{A}_x$, 
also leads to an interchange in the trifold of the two valleys. 
This interchange is equivalent to a reversal in the roles of the two valleys, consequently altering the sign of the integrated occupation difference by inverting the sign of these ratios [see Figs.~\ref{fig:MFIG_doccu_n2}(b) and (c)].

In the following we show the result for $p=-2$. The case for $p=4$ is summarized in the Appendices.

\section{Results}
\subsection*{Nanoribbon geometry}
Quasi-energy spectra of an irradiated graphene nanoribbon of finite length along the $x$ direction as a function of the lattice momentum $k_y$ for different values of $\phi$ are presented in Fig.~\ref{fig:qesdens}. 
As it is expected, $\phi = 0$ yields identical quasi-energy spectrum for both valleys. On the other hand, 
the spectrum becomes asymmetrical  for  $\phi=\pi/2$ 
and the minimas are located at $k_y\approx\pm 2\pi/3$ for both valleys with different quasi-energy values, 
represented by $\delta_{\pm}$ with $\delta_{+}<\delta_{-}$, 
as shown in Fig.~\ref{fig:qesdens}(b). 
The asymmetry reverses from one valley to another as $\phi$ transits from $\pi/2$ to $3\pi/2$ as evident from Fig.~\ref{fig:qesdens}(c).

\begin{figure}[t]
\centering
\includegraphics[width=0.98\linewidth]{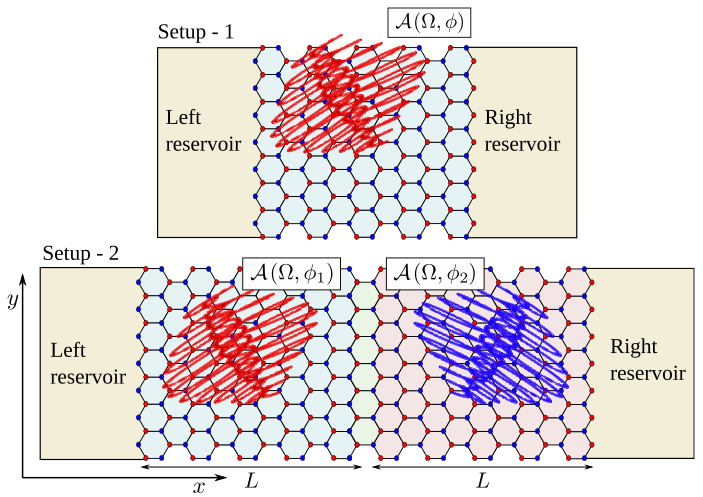}
\caption{Schematic of the device configurations. Top panel: a graphene nanoribbon connected to reservoirs at both ends is exposed to a bicircular laser field with total vector potential $\mathcal{A}(\phi)$. Bottom panel: Two graphene nanoribbons connected in sequence with their open ends linked to reservoirs are driven by the bicircular laser fields with vector potentials $\mathcal{A}(\phi_1)$ and $\mathcal{A}(\phi_2)$. We have considered the length of nanoribbons, denoted as $L$, which is transitionally invariant along the $y$ direction.} 
\label{fig:setup}
\end{figure}

In the following, we will analyse valley-sensitive  transport in two different physical setups. 
A single irradiated graphene nanoribbon is exposed to the bicircular field in setup 1 
and two reservoirs are connected to the left and right ends of the nanoribbon as shown in the top panel of Fig. \ref{fig:setup}.  
Without any relative bias of the reservoirs, the zero-temperature conductance is given as~\cite{KOHLER_2005,AK_2016}
\begin{equation}\label{eq:sigma_eq5}
\sigma(k_y) = \frac{e^2}{2\pi}\sum_{q\in\mathbb{Z}}\left[T^{(q)}_{\rm RL}(k_y,0) +T^{(q)}_{\rm LR}(k_y,0)\right],
\end{equation}
where $T^{(q)}_{\lambda\lambda'}(k_y,0)$ is the probability for an electron near Fermi energy to be transmitted from lead $\lambda'$ to $\lambda$ along with the absorption of $q$ photons with frequency $\Omega$. 
In this setup, let us introduce the polarisation due to valley asymmetry in the transport measurement as 
\begin{equation}\label{eq:VC1}
P_1 = \frac{\sigma_{+}-\sigma_{-}}{\sigma_{+}+\sigma_{-}},
\end{equation}
where $\sigma_{+(-)}=\sum_{k_y>(<)0}\sigma(k_y)$. Here, the contribution 
to the conductance for the two parts of the BZ is computed by restricting $k_y$  either to be positive or negative.

\begin{figure*}[ht!]
\centering
\includegraphics[width=0.9\linewidth]{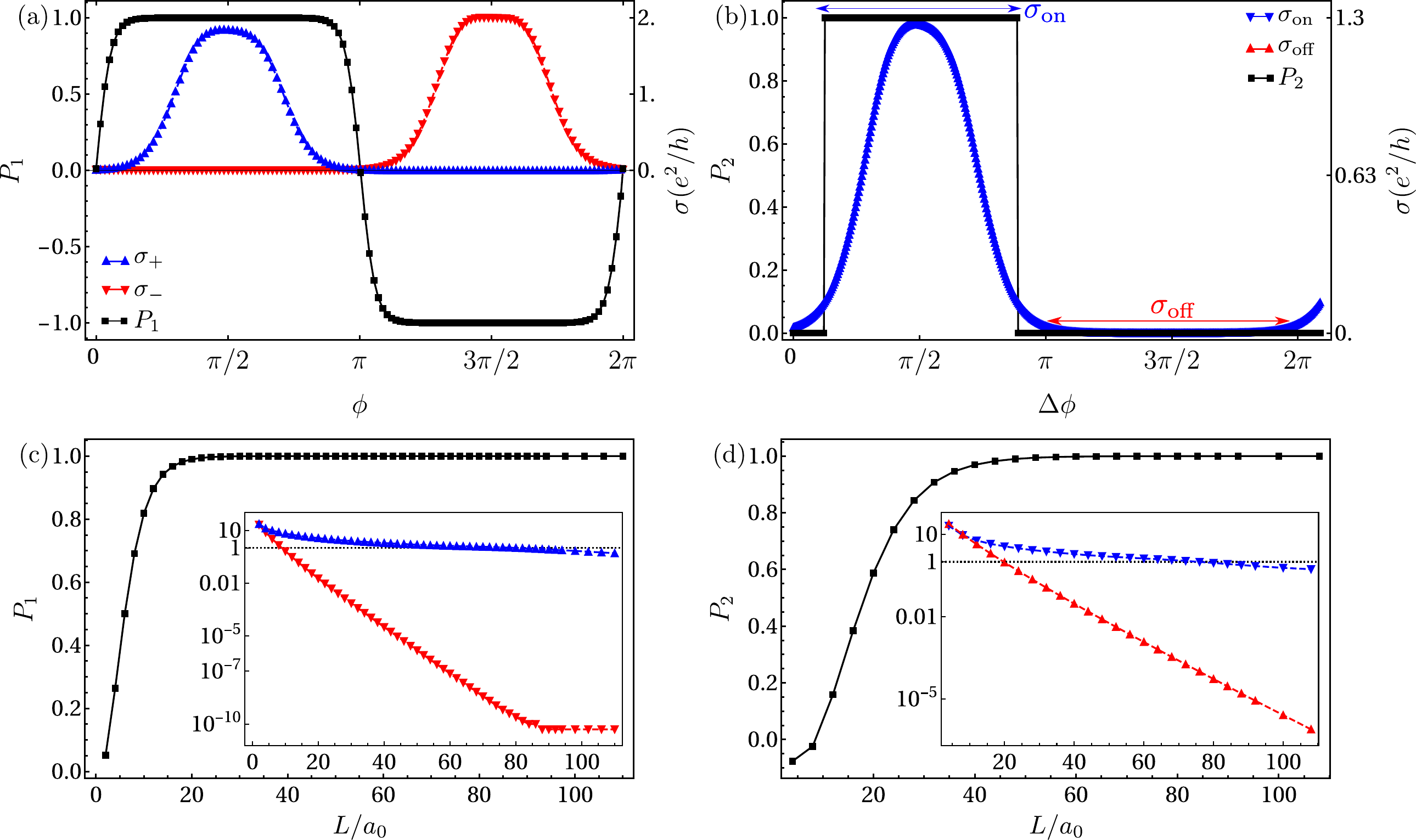}
\caption{For Configuration-1: Variations in the valley conductance ($\sigma_{\pm}$) and valley polarization ($P_1$)  
with respect to (a) the phase of the laser fields for nanoribbon length $L=40a_0$ and (c) the system size for $\phi=\pi/2$. 
For Configuration-2: Variations in the ``on/off" conductance ($\sigma_{\text{on/off}}$) and valley polarization ($P_2$)  with respect to (b) the phase of the laser fields for $L=64a_0$ and (d) the system size at phases 
$\Delta\phi_{\text{on}}  \Rightarrow \phi_1 = \phi_2 = \pi/2$ 
and $\Delta\phi_{\text{off}}  \Rightarrow  \phi_1 =  \pi/2$, and $\phi_2 = 3\pi/2$. Rest of the parameters are same as in Fig.~\ref{fig:n2hbz}.}
\label{fig:GNR1}
\end{figure*}

Figure~\ref{fig:GNR1}(a) presents 
variations in the polarisation  and valley conductances as a function of $\phi$ for an irradiated graphene nanoribbon in 
setup-1. 
The polarisation and  valley conductances can be periodically controlled by tuning $\phi$ as evident from the figure. 
This control stems from the fact that the emergence of the gapless states is intricately tied to $\phi$ as discussed above.
An absence of gapless states in both valleys results  a net zero conductance for $\phi=0$.  
As $\phi$ varies from 0 to $\pi/2$, only one valley possesses gapless states, 
leading to valley-polarized finite conductance output in that valley, while the conductance from the other valley remains  zero due to the absence of gapless states. This net conductance output is fully valley-polarized and on the order of $e^{2}/\hbar$. 
Changing  $\phi$ by $\pi$ reverses the situation with the roles of the two valleys interchanged. In this reversed scenario, the polarization becomes `$-1$', as the only nonzero contributions arise from the other valley as reflected from Fig.~\ref{fig:GNR1}(a).
The details of the momentum-resolved conductance contributions as a function of different parameters of the bicircular light is presented in the Appendix. 

Let us turn our discussion to another device configuration comprises of the sequential coupling of two  graphene nanoribbons, which are irradiated with two variations of the bicircular fields with vector potentials at phases $\phi_{1}$ and $\phi_{2}$ as depicted in the bottom panel of  Fig. \ref{fig:setup}.  
In this case, we define the relative phase as $\Delta\phi = \phi_{2} - \phi_{1}$, 
and the open edges of the system are connected to reservoirs, similar to setup-1. 

The polarization and the net conductance  as a function of $\Delta\phi$ with fixed $\phi_{1} = \pi/2$ for setup-2 is shown in Fig.~\ref{fig:GNR1}(b). 
The first half of the device configuration exhibits finite conductance and valley polarization with a polarization value of $P_1\approx1$ for $\phi_{1} = \pi/2$.  In this situation, the valley polarization of the second half becomes the governing factor in determining the overall net conductance output and 
the polarization of the second half can be fully controlled by varying the phase $\phi_2$. 
The net conductance can be computed as the sum of the conductances of these two halves in series. 
When the second half also exhibits the same polarization at $\phi_ 2 = 2n \pi+\pi/2$ with  $n$ as  an integer, 
both halves of the device conduct for the same valley, yielding a non-zero valley-polarized conductance output.

Another interesting scenario arises when $\phi_{2} = (2n+1)\pi+\pi/2$, which leads zero net conductance as 
the two halves of the system conduct for opposite valleys. 
We will henceforth refer to these phases as the ``on" and ``off" phases, respectively, as depicted in the figure. 
In these '``on" and ``off" phases, the phase difference between the two halves is determined by 
$\Delta\phi_{\text{on}}$ and $\Delta\phi_{\text{off}}$, given by $2n\pi$ and $(2n+1)\pi$, respectively.
We have quantified valley polarization in the valley valve configuration by comparing 
the relative conductance in the ``on" and ``off"  phases, denoted by $\sigma_{\text{on}}$ and $\sigma_{\text{off}}$ respectively, as $\sigma_{\text{on (off)} }=\sum_{k_y}\sigma(k_y)|_{\Delta\phi=\Delta\phi_{\text{on (off)}}}$.  Thus,  
the net valley-polarized conductance output can be quantify using the degree of valley polarisation as 
\begin{equation}\label{eq:VP2}
P_{2} = \frac{\sigma_{\text{on}}-\sigma_{\text{off}}}{\sigma_{\text{on}}+\sigma_{\text{off}}}.
\end{equation}
The key feature of the setup-2 is its ability to generate two distinct configurations, 
denoted as the ``on'' and ``off'' phases as evident from Fig.~\ref{fig:GNR1}(b).  
Therefore, device configuration in setup-2 can be seen as valley valve, which 
offers practical advantages for experimental applications due to its inherent ability to produce a valley-polarized net conductance output. 

\subsubsection*{Length dependence}
So far we have limited our discussion for a fixed length of the nanoribbon. It is important  to know how the 
nanoribbon's length influences the valley conductance and polarization for practical realisation of valley-based devices. 
The length-dependence of the valley conductances can be expressed as 
$\sigma_{\pm} \propto (w/L)\exp\left(-\delta_{\pm}L\right)$,  where $\delta_{\pm}$ represents the quasi-energy gaps of the two valley channels, $w$ stands for the width of the nanoribbon and $L$ denotes the length of the nanoribbon. 

\begin{figure*}[ht!]
\centering
\includegraphics[width=0.9\linewidth]{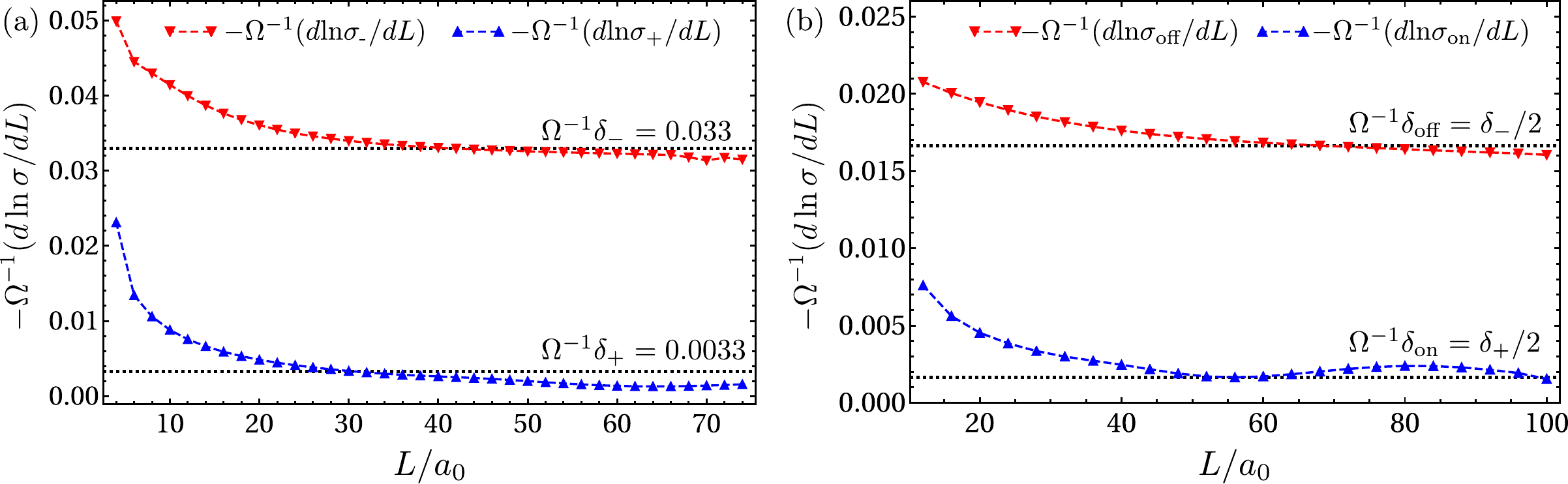}
\caption{(a) For Configuration-1: variations in quasi-energy gaps, $\delta_{\pm}=-d\ln\sigma_{\pm}/dL$ (in units of $1/\Omega$), as a function of $L$ (in units of $a_{0}$). 
The gridlines on the graph represent quasi-energy gaps observed in the quasi-energy spectrum shown in Fig.~\ref{fig:qesdens}. 
Rest of the parameters are same as in Fig.~\ref{fig:GNR1}(c). 
(b) For Configuration-2: variation in different quasi-energy gaps, $\delta_{\text{on/off}}=-d\ln\sigma_{\text{on/off}}/dL$ (in units of $1/\Omega$), as a function of $L$ (in units of $a_{0}$). The gridlines on the graph represent quasi-energy gaps $\delta_{\text{on/off}}$ computed from the analytical expression of the series sum of the conductances of the two halves of the irradiated graphene nanoribbons. Rest of the parameters are same as in Fig.~\ref{fig:GNR1}(d).}
\label{fig:GNR1_dlnsigma_vs_dL}
\end{figure*}

The sensitivity  of the valley conductance and polarization as a function of the length of the nanoribbon in setup-1 for 
$\phi=\pi/2$ is shown in  Fig.~\ref{fig:GNR1}(c). 
The dependence on length can be comprehended by considering the influence of quasi-energy gaps and the distribution 
of the Floquet states. In larger systems, one of the valley channels operates as a conductive pathway with a lower energy 
gap, while the other remains insulating with a more substantial gap [see Fig.~\ref{fig:qesdens}(b)].
The quasi-energy gaps can also be extracted from the functional dependence of conduction on length, which is 
defined as $(\delta_{\rm estimate})_{\pm}\approx -d\ln\sigma_{\pm}/dL$ when $L/a_0$ is large.
The extracted gaps are shown in Fig.~{\ref{fig:GNR1_dlnsigma_vs_dL}}, which 
are of the same order as shown in Fig.~\ref{fig:qesdens}(b) as the gaps computed from the quasi-energy spectrum.  
In sufficiently large systems, $\delta_{+}$ serves as the conductive channel, while $\delta_{-}$ maintains insulating characteristics. This regime, where $\delta_{+}<\delta_{-}$, results in perfectly valley-polarized transport.

The 	variation in ``on/off" conductance and 
valley polarization as functions of the nanoribbon length in setup-2 is presented 
in Fig.~\ref{fig:GNR1}(d). 
In this case, the conductance behavior  can be described as 
$\sigma_{\text{on (off)}}  \propto (w/L) \exp\left(-\delta_{\text{on (off)}}L\right)$, where 
$\delta_{\text{on (off)}}$ represents gap parameters controlling the length dependence of these conductances. 
These gap parameters can be estimated numerically as 
$\delta_{\text{on (off)}} \approx - d\ln\sigma_{\text{on (off)}}/dL$ at large $L/a_0$.
It is observed that these gap parameters, denoted as $\delta_{\text{on}}$ and $\delta_{\text{off}}$, can be approximated by the values $\delta_{+}/2$ and $\delta_{-}/2$, respectively as demonstrated in Fig.~\ref{fig:GNR1_dlnsigma_vs_dL}(b).

For a better understanding, we additionally computed the ``on/off"  conductances by summing the conductances of the 
two valley channels with distinct conducting and insulating behaviors interconnected in series. 
The conductance for the nanoribbon of length $L$ can be expressed as 
$\bar{\sigma}_{\text{on (off)}}  = \sigma_{+}\left(L/2\right)\oplus\sigma_{+ (-)}\left(L/2\right)$, where 
where $A\oplus B=\left(\frac{1}{A}+\frac{1}{B}\right)^{-1}$ and $\sigma_{+ (-)}\left(L/2\right)$ 
represents the conductance of the irradiated nanoribbon in steup-1 for the length $L/2$. 
The corresponding analytical expressions, derived from the analytical expressions of $\sigma_{\pm}$, can be simplified as follows
\begin{equation}\label{eq:Lconf2a}
\bar{\sigma}_{\text{on (off)}}  \propto \frac{(2) w}{L}\exp\left(-\frac{\delta_{+ (-)}L}{2}\right).
\end{equation}
From the above analysis, we observe that the conductances in setup-2 of length $L$ depend on the conductances of setup-1 with length $L/2$, resulting in a factor of $1/2$. 

\begin{figure}
\centering
\includegraphics[width=0.90\linewidth]{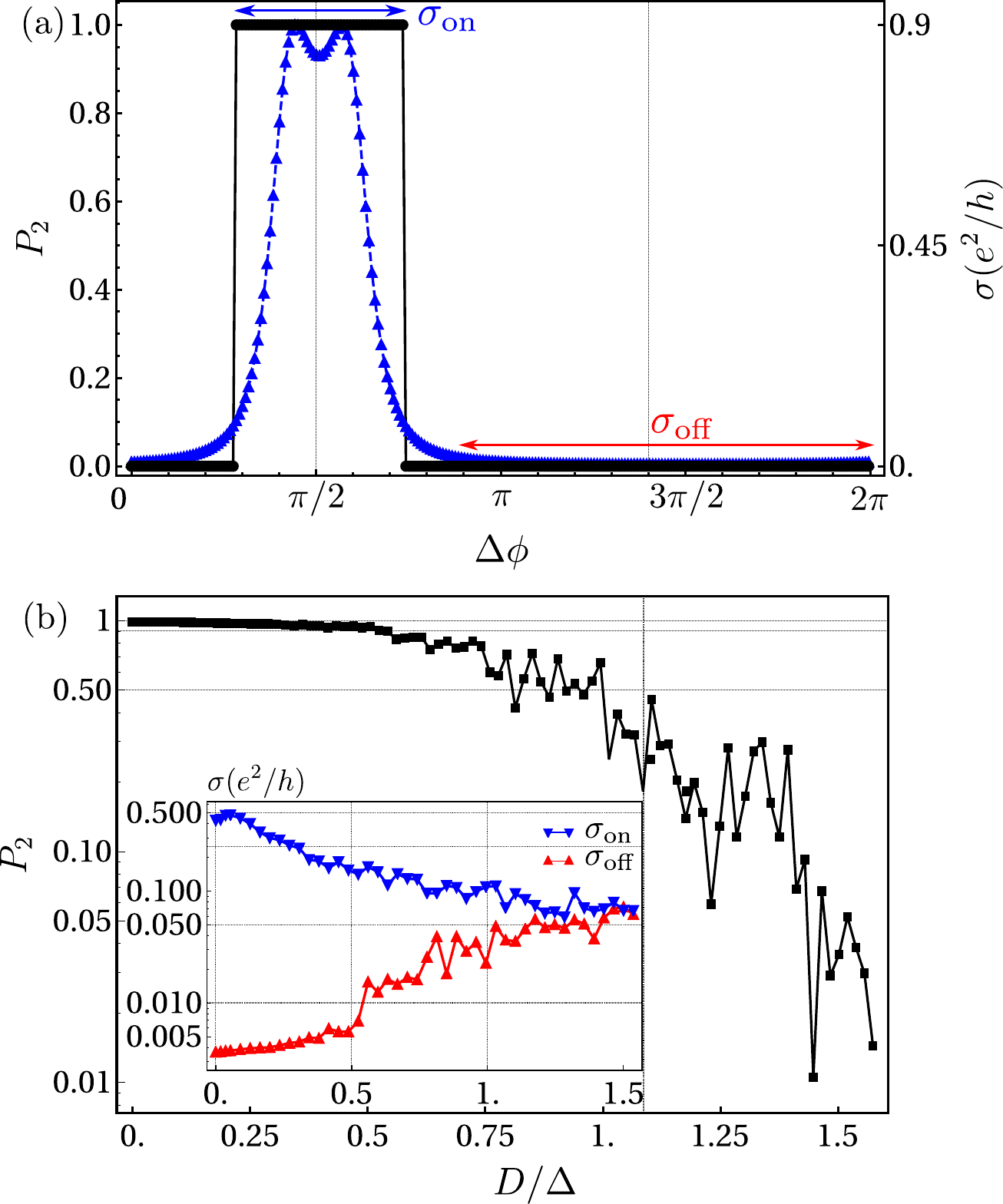}
\caption{For Configuration-2: (a) Variation in ``on/off" conductance and valley polarization as a function of the phase difference, and (b) with static disorder strength (in units of $\Delta=\delta_{-}-\delta_{+}\approxeq .03\Omega$) for the two-dimensional irradiated system with open boundaries along the $y$ direction. The parameters used are $L=40\,a_0$, $w=5\sqrt{3}a_0$, and the other parameters remain the same as in Fig.~\ref{fig:GNR1}.} \label{fig:MFIG_PvsD}
\end{figure}

\subsubsection*{Finite geometry and disorder dependence} 
Before concluding our findings, let us assess the robustness of our observed results in the presence of static disorder 
as the disorder is unavoidable in realistic situations. 
For this purpose, the ``on/off" conductances and polarization for a two-dimensional system with open boundary conditions are computed. 
In addition, the disorder is mimicked by the random static onsite potential,  which is represented by a Gaussian distribution with a zero mean and a standard deviation denoted as $D$ (in units of $\Delta$).  
All the  observables are averaged over random disorder realisations. 

The variation in ``on/off" conductance and polarization as function of the phase difference between the two halve of a 
two-dimensional system in the absence of any disorder is shown in Fig.~\ref{fig:MFIG_PvsD}(a) as a reference spectra.
Figure~[\ref{fig:MFIG_PvsD}](b) illustrates how the  ``on/off" conductance and polarization vary with the strength of the
static random disorder.
The presence of the static disorder diminishes valley polarization due to intervalley scattering, resulting in an increase in ``off" conductance. 
However, when the disorder strength remains moderate ($D/\Delta<1$), valley polarization retains its robustness. These findings indicate that disorder affects valley transport  but does not completely disrupt valley polarization as long as the disorder strength remains within certain limits.

\section{Discussion} 
In conclusion, present work unveils novel ways to leverage the valley polarization effect in pristine graphene exposed to a bicircular laser drive. 
We demonstrate that the bulk valley quasi-energy gaps can be fully controlled by manipulating 
the phase of the bicircular drive, thereby enabling precise control over valley occupations and valley conductance through tuning the drive's parameter. 
The proposed experimental setup for valley-filter and valley-valve devices provides a dynamic platform for capturing and tuning these valley polarization effects in transport experiments. 
Additionally, our findings reveal that these valley polarization effects remain robust even under moderate levels of static disorder. 
The proposed valleytronics devices offer a promising approach for optically manipulating valleys in graphene and similar systems, with potential applications in valley-based qubits, valley blockade devices, and other optically-enabled quantum electronic devices for quantum technologies.
		
\noindent{\it Acknowledgments:} R. K. acknowledges the use of PARAM Sanganak and HPC 2013, facility at IIT Kanpur. The support and resources provided by PARAM Sanganak under the National Super-computing Mission, Government of India, at the Indian Institute of Technology, Kanpur, are gratefully acknowledged. 
G. D. acknowledges support from Science and Engineering Research Board (SERB) India  (Project No. MTR/2021/000138). A.K. acknowledges support from the SERB (Govt. of India) via sanction No. CRG/2020/001803, DAE (Govt. of India) via sanction No. 58/20/15/2019-BRNS, as well as MHRD (Govt. of India) via sanction No. SPARC/2018-2019/P538/SL
		
\bibliography{new_refs_bib.bib}	
		
\clearpage
\onecolumngrid		
\begin{figure*}[ht!]
\centering
\includegraphics[width=.9\linewidth]{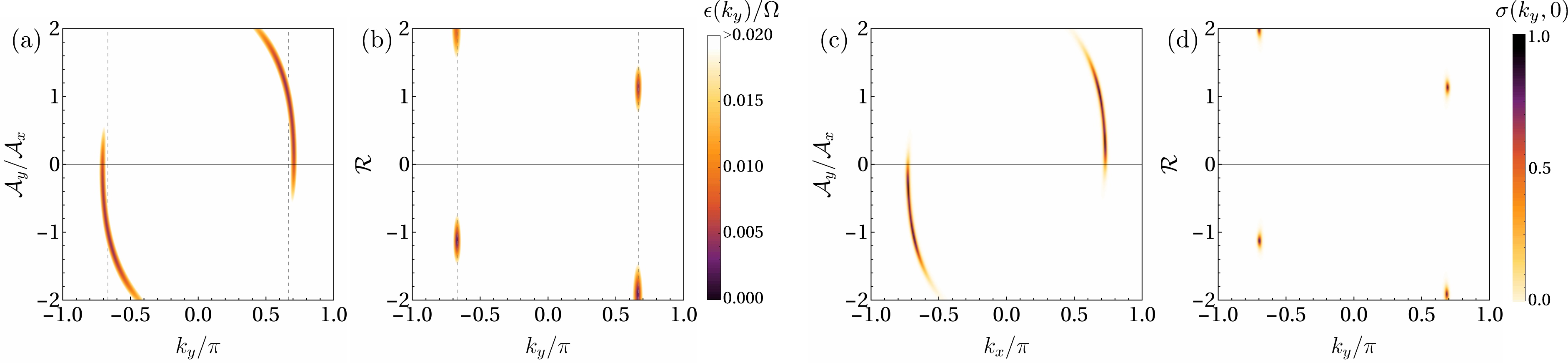}
\caption{Variation in the quasi-energy of a graphene nanoribbon with the periodic boundary condition as a function of  (a) $k_y/\pi$ and the ratio $\mathcal{A}_y/\mathcal{A}_x$, and (b) $k_y/\pi$ and the ratio $\mathcal{R}$.
(c)  and (d) are same as (a) and (b) for the variation in the conductance.  The other parameters  are $T=\gamma_0/4, L=40\,a_0$, $\phi=\pi/2$, $p =-2$, and $\mu =0$.}
\label{fig:Ratios}
\end{figure*}
\begin{figure*}[ht!]
	\centering
	\includegraphics[width=.95\linewidth]{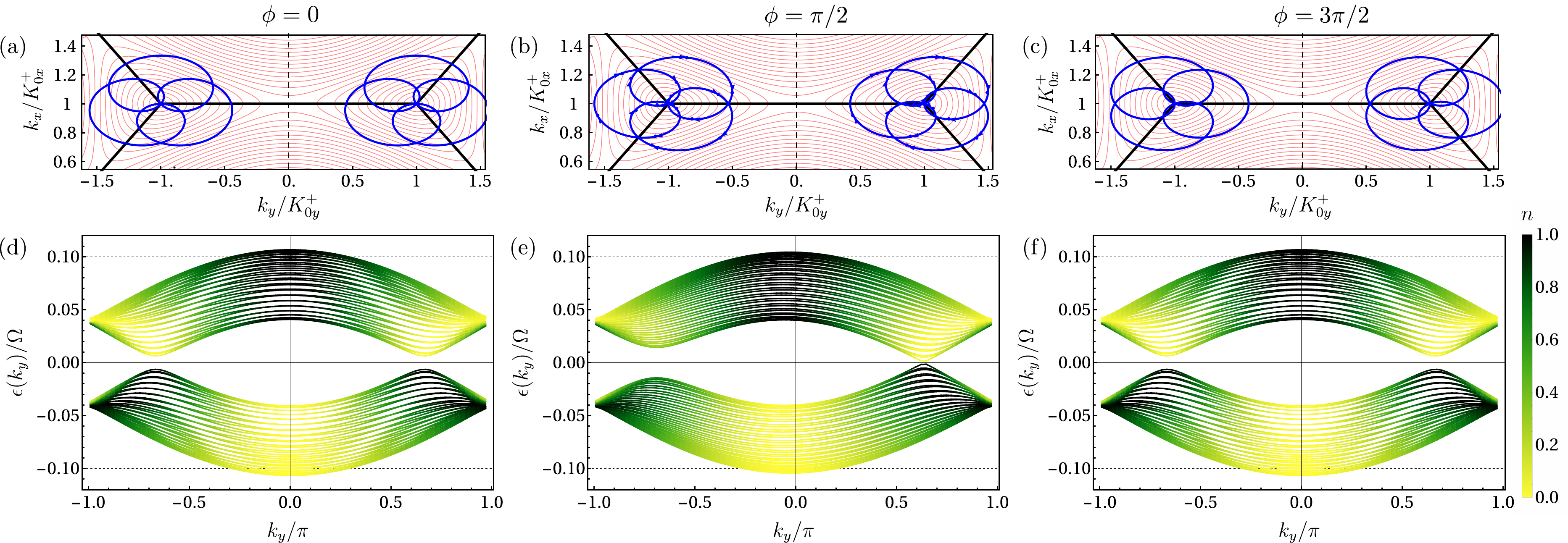}
	\caption{(Top) Projection of the Lissajous profiles  associated with the total vector potential of the bicircular counterrotating laser fields for  $p=4$ onto the low-energy contours of a pristine graphene near the Dirac points $K_{0}^{+}$ and $K_{0}^{-}$  for different values of the phase:  (a) $\phi=0$, (b) $\phi=\pi/2$, and (c) $\phi=3\pi/2$. The other parameters are $ \mathcal{A}_{x} = \mathcal{A}_{y} = 0.35$, $\mathcal{R}=-1$ and $T = 1$. (Bottom) Quasi-energy spectrum of  an irradiated graphene nanoribbon  exposed to  bicircular laser fields with periodic boundary conditions as a function of crystal momentum $k_y$ (in units of $1/\pi$) for (d) $\phi=0$, (e) $\phi=\pi/2$, and (f) $\phi=3\pi/2$. The colorbar represents the occupation of the quasi-energy states. The parameters are $p=4, T = \gamma_0/4, \mathcal{A}_0 = 1.75$, $\mathcal{R} = -1.07$ and $\mu = 0$.}
	\label{fig:n4hbz}
\end{figure*}
\twocolumngrid
\appendix	
\section{Momentum-resolved valley conductances:}
Fig.~\ref{fig:n4hbz}(a) shows the quasi-energy spectrum near the Fermi surface ($\mu=0$) as a function of $k_y$ and $\mathcal{A}_y/\mathcal{A}_x$, while (b) depicts the spectrum for $k_y$ and $\mathcal{R}$ at $\phi=\pi/2$. The figures demonstrate tunable quasi-energy gaps in the Brillouin zone by adjusting drive parameters. Corresponding two-terminal conductances are presented in Fig.~\ref{fig:n4hbz}(c) and Fig.~\ref{fig:n4hbz}(d), highlighting the role of quasi-energy states near the Fermi surface in the transport.
\begin{figure*}[ht!]
\centering
\includegraphics[width=0.9\linewidth]{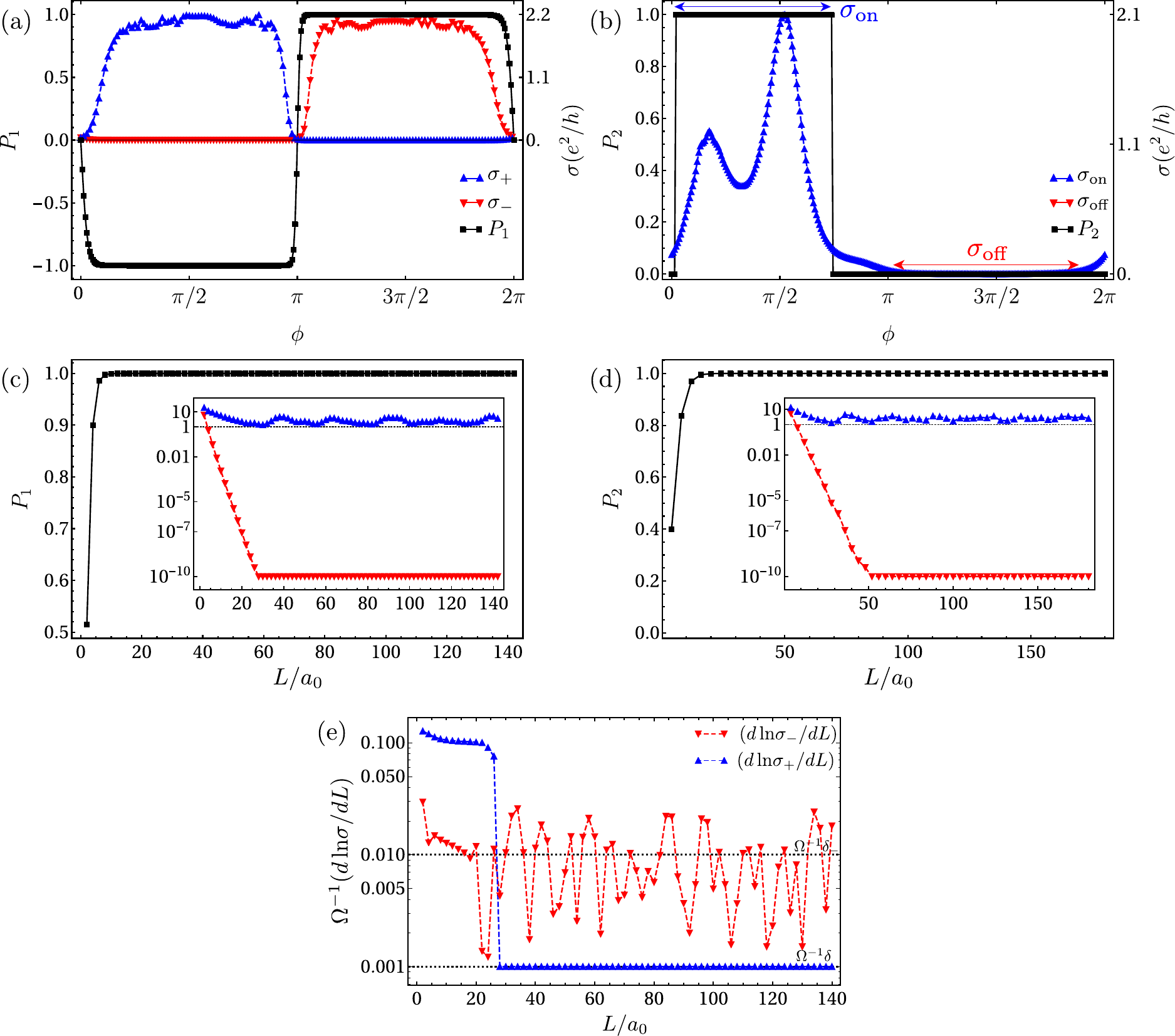}
\caption{For Configuration-1: Variations in the valley conductance ($\sigma_{\pm}$) and valley polarization ($P_1$)  
with respect to (a) the phase of the laser fields for nanoribbon length $L=20a_0$ and (c) the system size for $\phi=\pi/2$. 
For Configuration-2: Variations in the ``on/off" conductance ($\sigma_{\text{on/off}}$) and valley polarization ($P_2$)  with respect to (b) the phase of the laser fields for $L=20a_0$ and (d) the system size at phases $\phi_1 = \pi/2$, 
$\Delta\phi_{\text{on}} = 2n\pi$, and $\Delta\phi_{\text{off}} = (2n+1)\pi$. Rest of the Parameters are same as in Fig.~\ref{fig:n2hbz} with $\mu=.01$. (e) Variations in quasi-energy gaps, $\delta_{\pm}=d\ln\sigma_{\pm}/dL$ (in units of $1/\Omega$), as a function of nanoribbon length ($L$) for configuration-1. 
The gridlines on the graph represent quasi-energy gaps observed in the quasi-energy spectrum shown in Fig.~\ref{fig:n4hbz}.}
\label{fig:FIGS6}
\end{figure*}
\section{$p=4$ case}\label{sec:symmPm4}
The bicircular laser field with $p=4$ also  exhibits optical control over the valley symmetry of pristine graphene, similar to the previous scenario for $p = -2$. 
As illustrated in Fig.~\ref{fig:n4hbz}(Top row), the projection of Lissajous profiles onto the low-energy contours of pristine graphene illustrates the following observations: 
The laser field interacts identically with both valleys for $\phi=0$ as shown in Fig.~\ref{fig:n4hbz}(a). 
This situation changes as $\phi$ transits from $0$ to $\pi/2$. 
A complete alignment emerges between the Lissajous profile of the bicircular field with $\phi= \pi/2$ and the low-energy contours of one valley, while the alignment is entirely absent for the other valley as shown in Fig.~\ref{fig:n4hbz}(b). Notably, this alignment can be reversed by introducing a phase change of $\pi$ in this case as well. 
In contrast to $\phi=\pi/2$, $\phi=3\pi/2$ allows the roles of the valleys undergo a transition, and the trifolds Lissajous profile align with the other valley as shown in Fig.~\ref{fig:n4hbz}(c).
		
The quasi-energy spectrum of an irradiated graphene nanoribbon under the influence of a bicircular field with $p = 4$ 
is depicted in Fig.~\ref{fig:n4hbz}(Bottom row). 
In this scenario, the periodic modulation of the optical inversion symmetry breaking is demonstrated. 
For $\phi=0$, the quasi-energy spectrum of both valleys is identical and exhibits a finite energy gap as illustrated in Fig.~\ref{fig:n4hbz}(d). 
Upon changing the phase to $\phi=\pi/2$, the quasi-energy spectrum undergoes a transition, with one valley displaying gapless states while the other remains gapped as shown in Fig.~\ref{fig:n4hbz}(e). 
Conversely, when the phase is tuned to $\phi=3\pi/2$, the situation observed in Fig.~\ref{fig:n4hbz}(f) is reversed. Gapless states emerge in the valley where they were previously absent, in accordance with the alignment of Lissajous profiles with the low-energy contours of pristine graphene. 
Thus, the periodic modulation of the optical inversion symmetry breaking with $\phi$ is visible from  Fig.~\ref{fig:n4hbz}. 

In the case of $p=4$, results for two-terminal transport show similar conductance behavior as observed previously for  $p=-2$ [see  Fig.~\ref{fig:FIGS6}]. 
We find that the conductance in the valley region exhibits periodic behavior with respect to $\phi$. 
When $0 \leq \phi < \pi$, the contribution to the conductance arises predominantly from one valley as the other valley exhibits finite energy gaps, as evident in Fig.~\ref{fig:n4hbz}(d-f). 
However, the alternative valley begins to contribute to the overall conductance for $\pi \leq \phi < 2\pi$. 
These contributions are quantified as $\sigma_{+}$ and $\sigma_{-}$ for the respective valleys as shown in Fig.~\ref{fig:FIGS6}(a).

It is feasible to modulate the conductance in an ``on/off" fashion by controlling the phases of the two halves of the system 
in configuration-2 setup. 
For $\phi_1 = \pi/2$ and $\phi_2 = \phi_1 + 2n\pi$,  the system operates in the ``on-phase", resulting in a finite conductance output as both halves of the system are conductive for the same valleys. 
Interestingly, the system can be tuned to the ``off-phase" for $\phi_2 = \phi_1 + n\pi$, 
where the net conductance becomes zero, as both halves of the system become conductive for opposite valleys.

We have also explored the conductance characteristics as a function of the nanoribbon's length 
for configuration-1 and configuration-2 set ups as shown in Fig.~\ref{fig:FIGS6}(c) and Fig.~\ref{fig:FIGS6}(d), respectively. 
In both cases, we observe that the functional dependencies of the conductances can be expressed in terms of the expression as discussed in the main text for $p=2$. The gaps $\delta_{\pm}$  computed  with respect to $\mu$ is  shown in Fig.~\ref{fig:FIGS6}(e).

\end{document}